%% file: paper.tex
\documentclass[aps,prd,twocolumn,floatfix,superscriptaddress]{revtex4}
\usepackage{amsmath}

\usepackage[dvips]{graphicx}

\newlength{\colw}
\begin{document}

\title{Radial and orbital excitations of static-light mesons}
\author{Justin Foley} 
\affiliation{Department of Physics, Swansea University, Singleton Park, 
  Swansea SA2 8PP, UK}
\author{Alan \'{O} Cais} 
\altaffiliation[Current address: ]{CSSM, School of Chemistry and Physics,
University of Adelaide, Adelaide SA5000, Australia} 
\author{Mike Peardon} 
\affiliation{School of Mathematics, Trinity College, Dublin 2,
  Ireland}
\author{Sin\'{e}ad M. Ryan}
\affiliation{School of Mathematics, Trinity College, Dublin 2,
  Ireland}
\date{\today}
\preprint{TrinLat-07/02}
\begin{abstract}
We present results for the spectrum of static-light mesons from 
$\rm{N}_{\rm{f}} = 2$ lattice QCD. These results were obtained using 
all-to-all light quark propagators on an anisotropic lattice, yielding an 
improved signal resolution when compared to more conventional lattice 
techniques. With a light quark mass close to the strange quark, we have measured the splittings 
between the ground-state S-wave static-light meson and higher excitations.
We attempt to identify the quantum numbers of the excited states in the context of the reduced spatial 
symmetries of the lattice. 
\end{abstract}

\maketitle

\input{intro}
\input{sl-spectrum}

\input{grouptheory}

\input{latticeoperators}

\input{variational}
\input{all-to-all}

\input{latticeactions}
\input{simulationdetails}

\input{results}

\input{finitevolume}

\input{massandspacing}
\input{multiparticle}
\input{discussion}

\input{conclusions}
\bibliography{paper}

\end{document}

%% file: intro.tex
\section{Introduction}
\label{intro}
The physics of heavy-light hadrons plays a crucial role in the
determination of CKM matrix elements to test the Standard Model and
search for new physics. In particular, systems with one or more bottom 
quarks are of greatest interest but are also the most theoretically
and technically challenging. Recently, experiments have begun to probe
the excited state spectrum of $B$ mesons and signals for a number of
new states have been seen. It is still unclear if these new
states have the expected masses and energies or
if they are completely new states, not predicted by the standard
theory. 

Lattice QCD, in principle, offers the
possibility of {\it ab initio} calculations of the relevant hadronic
parameters. However, mass-dependent errors which arise in the
discretisation of the Dirac operator have limited the accuracy of
heavy-light simulations. Effective theories, which can be used to control or 
eliminate these errors have been developed and widely used in numerical
simulations to predict the masses and decay form factors of stable ground
states to better than 10\% precision~\cite{Onogi:2006km}. 
The static quark approximation is the starting point for
 one such approach. It is based on the observation that to a
good approximation a heavy-light hadron can be described by an idealised 
system in which the heavy quark is infinitely massive. Heavy-light
hadrons exhibit approximate heavy quark spin and flavour symmetries
which arise because the Compton wavelength of the heavy quark is much smaller 
than the typical hadronic length scale ($m_{Q} \gg
\Lambda_{QCD}$). In the static limit these symmetries are exact and
the hadron consists of light degrees of freedom bound to a static
colour source. Systematic corrections to the static limit can be 
computed order-by-order in $\Lambda_{QCD}/m_{Q}$, which forms the 
basis for heavy quark effective theory (HQET). This approach is 
particularly suitable for hadrons containing a bottom quark where 
leading order corrections to the static limit, which include 
chromomagnetic interactions and the heavy quark kinetic energy, are 
at the ten-percent level. Although the static approximation is no
longer valid for the much lighter charm quark the approach is
nevertheless useful. Relativistic simulations at quark masses below
charm can be combined with a simulation in the static approximation to
allow interpolation to the charm quark mass and in fact this method can
also be used to determine $b$ physics. 

Given the utility of the static-quark approximation it may at first 
appear surprising that this formalism is so under-utilised in lattice 
simulations. However, simulations in the static quark approximation 
have suffered from one significant drawback -- the signal-to-noise
ratio is very poor. Ultimately, this is due to the fact that a static 
quark propagates only in time. While this makes the static quark propagator 
trivial to evaluate, in a conventional lattice simulation where
one computes only the elements of the light-quark propagator from a single 
space-time site to all other lattice sites, information on static
quark propagation is only obtained from a single spatial lattice site.

Efforts to overcome this problem have focused on two areas. 
One approach is based on the observation that the signal-to-noise
ratio of a static-light two-point function decays exponentially in time
with an exponent which receives a large contribution from the
static quark self energy. The Alpha collaboration realised that
by carefully choosing the discretisation of the static quark
action it is possible to reduce the exponent, thereby improving
the signal resolution at large times~\cite{DellaMorte:2005yc}.

The second approach, which is used here, is to stochastically estimate 
the full Dirac
propagator~\cite{Bitar:1988bb,Kuramashi:1993ka,Dong:1993pk,deDivitiis:1996qx,Eicker:1996gk,Michael:1998sg,McNeile:2000xx,Wilcox:1999ab,
Neff:2001zr,Duncan:2001ta,DeGrand:2002gm,Bali:2005fu,Peardon:2002ye,Foley:2005ac}.
This allows source and sink operators for the static-light correlators 
to be placed at every spatial lattice site, yielding a dramatic
increase in statistics. 
In this case, it is crucial to optimise the propagator estimate
to avoid the introduction of unnecessary noise. In addition to the 
vast gain in statistics, the use of all-to-all propagators allows 
for complete freedom in the choice of interpolating operators
since the measurement process is decoupled from the evaluation of 
the propagator. Once the propagator estimate has been
computed, one can construct any spatially-extended interpolating field
or apply any level of smearing to the quark fields  at both the source
and sink, without the
need for additional fermion matrix inversions. Therefore, an
efficient, practical all-to-all algorithm~\cite{Foley:2005ac} 
allows us to access not only
ground state resonances, but also excited hadronic states.

Of course, the two improvements can be combined and a simulation
incorporating both the improved static action and all-to-all
propagators may prove to be even more precise. This has not yet been
tested. 

This paper is concerned with a precision determination of 
the excited-state spectrum 
of the $B_s$ meson. The spectroscopy of the
stable ground-state heavy-light meson is already well-studied on the
lattice using a variety of approaches including the static
approximation~\cite{Eichten:1989kb}, the Fermilab~\cite{El-Khadra:1996mp} approach and NRQCD~\cite{Caswell:1985ui}. However, accurate
simulations of the spectrum of radial and orbital excitations, in
practice, require an all-to-all propagator algorithm and as a 
consequence are not as mature.

In this study, we compute the excited-state spectrum of static-light
mesons in two-flavour QCD. Experimentally, little is
known of the excited-state spectrum of heavy-light hadrons and so 
predictions from lattice QCD are of considerable 
phenomenological interest. Simulations of static-light mesons can also
be seen as a natural testing ground for all-to-all propagator techniques 
since they require only a single light propagator per configuration. 
It is therefore unsurprising that a number of groups have attempted
to tackle this particular problem. The static-light excited-state spectrum
in dynamical QCD has previously been studied in 
refs.~\cite{Koponen:xxx,Burch:2006mb}.

In this paper we compute the energies of static-light mesons 
where the light quark mass is close to the strange quark. Our results 
are therefore most relevant for the $B_s$ spectrum. In this case, 
experimentally, the ground state (pseudoscalar and vector) energies
are well established, but there exists no conclusive experimental
determination of the excited states~\cite{Yao:2006px} \footnote{
The quantum numbers of the $B_{s J}^{*}(5850)$ are as yet unknown.}.

The remainder of the paper is organised as follows. 
Section~\ref{staticlight-spectrum} describes the group theory of the 
static-light spectrum in the continuum and its analogue on the
lattice. The construction of lattice operators, their analysis using 
a variational approach, the all-to-all propagators and lattice actions 
used are also described. Section~\ref{results} discusses the results 
of the dynamical simulations including finite volume effects and the 
spectrum of states. In Section~\ref{multiparticle} we address the
issue of multiparticle states. 
Section~\ref{discussion} contains a discussion of 
the results and in Section~\ref{conclusions} we draw some conclusions. 

%% file: sl-spectrum.tex
\section{The static-light spectrum}
\label{staticlight-spectrum}
The heavy-quark spin symmetry in the static limit means that mesons which
differ only by the spin of the heavy quark are degenerate, ie,
there is no hyperfine splitting. In this limit
the total angular momentum and parity of the
light degrees of freedom ($J_{\ell}^{P_{\ell}}$) are conserved quantities, and
it is conventional to use these quantum numbers to label the
degenerate hyperfine multiplets~\cite{Isgur:1991wq}.
So, for example, the pseudoscalar ($0^{-}$)
and vector  ($1^{-}$) mesons correspond to 
$J_{\ell}^{P_{\ell}} = {\frac{1} {2}}^{-}$.
In the constituent quark model description of hadrons these
are S-wave (orbital angular momentum L=0) channels.
Similarly the P-wave channels have 
$J_{\ell}^{P_{\ell}} = {\frac{1} {2}}^{+}, {\frac{3} {2}}^{+}$
and the D waves are 
$J_{\ell}^{P_{\ell}} = {\frac{3}{2}}^{-}, {\frac{5} {2}}^{-}$.
Note that we have implicitly assumed that the mesons contain a static 
quark, which has positive intrinsic parity.
In our measurements we exploit the heavy-quark
spin symmetry to optimise our signals
and average over all degenerate channels.
To facilitate this we construct interpolating operators
specifically for the light degrees of freedom.
We then combine the source and sink operators with a
simple Wilson line to obtain a gauge-invariant two-point
function with the required quantum numbers~\cite{Eichten:1989kb}.

A lattice determination of static-light energies is important in its
own right but in addition, such an unambiguous determination of the
spectrum can also be used to indicate which quark models most
accurately reproduce the static-light physics. Naively, for a given 
value of L, one expects a natural ordering of energies to prevail.
That is, the multiplet with a larger value of
$J_{\ell}$ has a higher energy. However, some 
quark models predict that at higher
orbital and radial excitations this ordering
should reverse~\cite{Schnitzer:1989xr,Isgur:1998kr}.
Whether this reversal in ordering
does in fact occur, and at what level it occurs, 
is still a matter of debate. This
phenomenon was first investigated in a dynamical lattice simulation in ref.~\cite{Green:2003zz}
which concluded that no reversal of ordering occurs
up to and including D-wave excitations.

%% file: grouptheory.tex
\subsection{Spatial rotations on the lattice}
\label{grouptheory}
The accurate determination of the excited-state
spectrum requires a precise understanding of lattice space-time
symmetries.  The angular momentum quantum number $J$, assigned to 
states in the continuum, labels irreducible representations (irreps) 
of $SO(3)$, the group of proper rotations which leaves the continuum
QCD Hamiltonian invariant. On a lattice which is isotropic in space, 
this symmetry group is broken to a 24-element subgroup - the octahedral 
point group $O$.

If we base our analysis solely on the continuum symmetry group
then we will almost certainly misidentify
some of the excited-state resonances.
We therefore compute correlation functions of operators
which transform irreducibly under the lattice symmetry group.
These operators have non-zero overlap with an infinite
number of states, and the continuum quantum numbers
of the low-lying states can be deduced by observing (near) 
degeneracies across the lattice irreps.

The full spatial symmetry group is in fact $O_{h} = O \otimes C_{2}$, 
where $C_{2}$ consists of the identity and the space inversion operator.
The elements of $C_{2}$ commute with the group of proper rotations and
the irreps of the full symmetry group follow
directly from the irreps  of $O$. Therefore we first consider the
irreps of this subgroup.

The octahedral point group has five conjugacy
classes and accordingly five single-valued irreps.
These are labeled $A_{1}$, $A_{2}$, $E$, $T_{1}$ and $T_{2}$,
and are of dimension 1,1,2,3 and 3 respectively.
Determining the relationship between the angular momentum quantum
number and these irreps is relatively straightforward.
Restricting the continuum irreps, labeled by $J$, to the elements
of $O$ generates representations for $O$ which are in general
reducible. These \textit{subduced} representations can then be decomposed
into their constituent irreps.

The number of times a particular irreducible representation
of $O$, labeled by the superscript $\alpha$, occurs
in the continuum $J$ irrep is given by the formula
\begin{eqnarray}
n_{J}^{(\alpha)} = \frac{1} {N_{G}} \sum_{k} n_{k} \chi_{k}^{(\alpha)} \chi_{k}^{(J)}.
\label{eqn:count_irreps}
\end{eqnarray}
$N_{G}$ is the number of group elements, in this case $N_{G} = 24$.
The subscript $k$ labels the conjugacy classes of the lattice
symmetry group, and $n_{k}$ is the number of
group elements in a conjugacy class. $\chi_{k}^{(\alpha)}$ is the character
of the conjugacy class in the lattice irrep and $\chi_{k}^{(J)}$
is the corresponding character in the subduced representation.
By applying this formula systematically for a number of values
of $J$, it is possible to deduce the angular momentum content of the
lattice irreps.

The preceding analysis applies to single-valued, or bosonic, 
representations only. However, interpolating operators
for the light fermionic degrees of freedom transform according
to the double-valued irreps of $O$. To find these, we
consider the double-cover group $O^{D}$. This 48-element
group has eight irreps, but five of these coincide with the
single-valued irreps of $O$. The remaining three are labeled
$G_{1}$, $G_{2}$ and $H$, and these form the double-valued irreps
of $O$.
$G_{1}$ and $G_{2}$ are both two-dimensional
representations, and $H$ has dimension four.
The angular-momentum content of these irreps
can be determined using the formula in Eq.(\ref{eqn:count_irreps}),
where in this case the sum extends over the elements of $O^{D}$, 
for which $G_{1}$, $G_{2}$ and $H$ are single-valued representations.

The extension of this discussion to the
 full spatial symmetry group is clear: $O_{h}$ has twice the number of
irreps as $O$. For example, the irrep $G_{1}$ of $O$ yields the irreps
$G_{1 g}$ and $G_{1 u}$ of $O_{h}$. The subscripts $g$ and $u$ denote 
representations which are even (gerade) and odd
(ungerade) under spatial inversion, respectively.
Table~\ref{tab:irreps-of-Oh} lists the even-parity irreps of $O_{h}$ and the quantum numbers
of their low-lying constituent states. 
\begin{table}[h]
\begin{center}
\begin{tabular}{ccc}
\hline
Lattice irrep & Dimension & ${J}^{P}$ \\
\hline
${{G}}_{1 g}$ & 2 & $\frac{1} {2}^{+}, \frac{7} {2}^{+}$... \\
${{G}}_{2 g}$ & 2 & $\frac{5} {2}^{+}, \frac{7} {2}^{+}$... \\
${H}_{g}$ & 4 & $ \frac{3} {2}^{+}, \frac{5} {2}^{+}, \frac{7} {2}^{+} $..  \\
\hline
\end{tabular}
\end{center}
\caption{Even-parity irreducible representations of the octahedral point group with 
the quantum numbers of the lowest-lying constituent states.}
\label{tab:irreps-of-Oh}
\end{table}
The table illustrates an obvious but important point - states which lie in the
same continuum irrep, but with different values of $J_{z}$ (labeling the rows
of the continuum irreps) are in general divided between the lattice irreps.
For example, states with quantum numbers $J_{\ell}^{P_{\ell}} = {\frac{5} {2}}^{+}$,
 corresponding to a 6-dimensional representation of $O(3)$, appear in both the $G_{2 g}$ and
$H_{g}$ irreps. In the continuum, in the absence of an external field, these states are degenerate
in energy; however, this degeneracy is broken by lattice artifacts. Therefore, in a numerical
study one hopes to determine the ${\frac{5} {2}}^{+}$ energy levels by identifying near-degenerate
levels in the $G_{2 g}$ and $H_{g}$ irreps which converge in the approach to the continuum
limit.

We are now in a position to assign the states of interest
to their respective lattice irreps.
We see that the S wave, which has
$J_{\ell}^{P_{\ell}} = {\frac{1} {2}}^{-}$, lies in the
$G_{1 u}$ irrep.
The ${\frac{1} {2}}^{+}$ P wave and the
${\frac{3} {2}}^{+}$ P wave appear in the
$G_{1 g}$ and $H_{g}$ irreps respectively.
The D-wave multiplets are labeled ${\frac{3} {2}}^{-}$
and ${\frac{5} {2}}^{-}$.
Both of these appear in the $H_{u}$ irrep, and the
${\frac{5} {2}}^{-}$ states also arise in the $G_{2 u}$ irrep.
The lowest-lying states in the remaining lattice irrep , the
$G_{2 g}$ representation, are expected to have the quantum numbers
${\frac{5}{2}}^{+}$, corresponding to F-wave excitations.

%% file: latticeoperators.tex
\subsection{Lattice operators}
\label{latticeoperators}
To construct operators which transform
according to the irreps of $O_{h}$,
we first
identify sets of linearly-independent prototype operators which
transform amongst themselves under $O_{h}$.
The static quark is fixed in space and in this discussion it is
convenient to identify the position of the static quark with
the origin of a coordinate system. Suitable prototype operators
then consist of the spin components of the light quark field
which can sit at the origin, yielding local operators, or may be separated
from the static quark by a gauge-covariant product of link variables.

The action of $O_{h}$ on these prototype operators generates a
representation of the group, the operators are said to
transform according to the
rows of that representation.
Interpolating operators which transform according to the
constituent irreps can be obtained by taking linear combinations
of the prototype operators.
A detailed description of how one can construct a complete
set of basis operators from a prototype set
for the more complicated case of baryon spectroscopy is given in ref.~\cite{Basak:2005aq}.
Here, we simply note that by construction,
basis operators for a given instance of a lattice irrep
form an orthogonal set. 

Before we proceed, it is worth clarifying our
terminology.
Obviously, by choosing different sets of prototype operators it is possible
to construct a number of basis sets for a given irrep,
and it may also happen that a particular irrep appears
more than once in the decomposition of one of the prototype
representations. 
For the remainder of the paper, we will say that each disjoint set,
transforming according to a particular irreducible representation, corresponds
to a single
 `instance' of that representation.

We work in a spin basis where the upper
components of a Dirac spinor have positive parity, while the lower
components have negative parity.
We can therefore restrict ourselves
to prototype operators which use only the upper or lower
components of the light quark fields.
Operators appearing in a given instance of a representation
will only contain either the upper or lower quark field spin components,
and we note that exchanging the
upper and lower spin components
simply changes the parity of the operators.

\begin{figure}[h]
\centering
\resizebox{0.48\textwidth}{!}{
	\includegraphics*{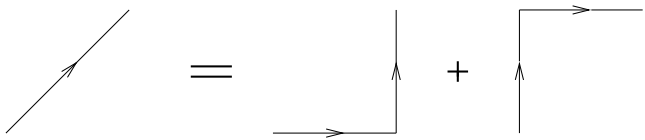}
}
	\caption{The planar diagonal path connecting the static quark 
to the light quark field components, which is required to construct 
operators for  $G_{2 u}$, is given by the sum of two L-shaped 
paths.}
      \label{fig:diagonal}	
\end{figure}

First we consider local operators.
Under $O_{h}$ the upper components of the quark field
transform according to the $G_{1 g}$ irrep, while the lower
components lie in $G_{1 u}$. All other irreps require
spatially-extended operators. The simplest spatially-extended
prototype operators consist of light quark field components
separated from the static quark by straight-line paths of equal length.
For a given path length, 6 offsets need to be considered -
displacements along the $x$, $y$ and $z$ axes in positive and negative
directions. Combining these offsets with two of the four light
quark field spin components yields a 12-dimensional representation.
This representation decomposes into $G_{1 g}$, $G_{1 u}$, $H_{g}$ and
$H_{u}$. All the S, P and D-wave states appear in these irreps.
The four irreps can be studied using only local operators  
and operators with straight-line offsets. 
However, identification of the D-wave states also requires the use of operators
which project onto the $G_{2 u}$ irrep. The simplest such operators
can be obtained from a prototype set consisting of spin components 
displaced from the origin along diagonals in planes spanned by the 
$x$, $y$ and $z$ axes. In terms of link variables these  
diagonals are given by the sum of two L-shaped paths as shown 
in Fig.~\ref{fig:diagonal}.
We can consider prototype operators with
more complicated paths to try to identify operators which better overlap
with the states of interest. However, for our purposes, operators
involving these simplest paths yielded good results.
In practice, we do not use `bare' links and quark fields
in these interpolating operators. In order to reduce
noise in our measurements, the link variables are stout-smeared~\cite{Morningstar:2003gk,Basak:2005gi}.
In addition, as we describe later, Jacobi smearing~\cite{Gusken:1989ad} of the light-quark fields
plays a central role in the determination of the excited-state energies.
In both cases the smearing is implemented such that the smeared
quark fields and link variables have the same transformation
properties under $O_{h}$ as their unsmeared counterparts; therefore, 
the statements made above about the choices of path required to access
each irrep are unchanged.

Within this collection of operators, there is still some freedom
in the choice of interpolating operators.
This is particularly evident in the case of straight-line displaced
operators, where the prototype representations
using either the upper or lower components
have the same decomposition in terms of the lattice irreps.
From the operator set, final measurements are performed with the 
operators which couple optimally to the states of interest. The 
correlation functions we wish to evaluate
give the amplitude for creating a static-light
meson, containing a static quark, with the required quantum numbers at
some initial time and annihilating the meson at a later time.
When we evaluate the amplitude for this process we find that 
interpolating operators which use the upper components of the
light quark field yield noisy measurements. Fortunately, the operators
which use the lower quark field components have a larger
overlap with the states of interest and give much cleaner
signals. To avoid the introduction of unnecessary
noise, we restrict ourselves to operators containing the lower
quark field components in the evaluation of two-point functions which 
describe a meson which contains a static quark, propagating forwards in time. 

Therefore, the only local operators used transform
according to the $G_{1 u}$ irrep, and we use spatially-extended
operators to project onto the $G_{1 g}$ representation.
As  discussed, operators with straight-line paths
project onto the $H_{g}$ and $H_{u}$ irreps, and it is possible
to construct operators incorporating planar-diagonal
paths which project onto the $G_{2u}$ irrep.
Similar operators can be used to project onto the 
$G_{2 g}$ irrep.
However, the lowest-lying states in this representation
are the $\frac{5} {2}^{+}$ and $\frac{7} {2}^{+}$ F waves. 
These very heavy states
are not of immediate interest to us, and we do not
include the $G_{2 g}$ irrep in this study.

Therefore it is possible to evaluate correlation functions for
five lattice irreps using only the lower quark field components and
the simple gauge-covariant paths listed previously.
However, using the transformation property of the fermionic degrees of freedom 
under time reversal, the upper spin components of the light quark field
can be used to further enhance our statistics. 
So far, the correlation functions have been described in
terms of a meson containing a static quark, propagating forwards in time.
Formally, the same amplitude can be obtained by reversing
the direction of the Wilson line and substituting the upper quark field
spin components into the interpolating operators.
This is a correlation function for the light degrees of freedom
with the same angular momentum, but opposite parity,
propagating backwards in time. It can also be
regarded as a correlation function for a static-light
meson containing a static anti-quark.
By averaging over correlators for mesons containing a static
quark, and mesons containing a static anti-quark, we ensure that
all four light quark spin components are used in our measurements.

%% file: variational.tex
\subsection{The variational method}
\label{variational}
The operators described in the previous section
couple to all the states appearing in their respective irreps.
At sufficiently large times it is expected that the
behaviour of the corresponding two-point functions will
reduce to the usual single exponential decay with
an exponent equal to the energy of the lightest state
appearing in that irrep. A variational approach must therefore
be applied to
compute the energies of higher-lying
states in each irrep.
To apply the variational method in the lattice irrep
$R$, we first compute a matrix of correlation functions
\begin{eqnarray}
C_{\alpha \beta}(\tau) &\propto&
 \sum_{t, \vec{x}} \langle \Omega |
\overline{\mathcal{O}}_{\alpha}^{(R)}(t+\tau, \vec{x}) \mathcal{W}(t+\tau,t;\vec{x})
\mathcal{O}^{(R)}_{\beta}(t,\vec{x}) | \Omega \rangle \nonumber \\
& & + \rm{T.R}
\label{eqn:corr_matrix}
\end{eqnarray}
where the subscripts $\alpha, \beta$ label
interpolating operators for the light degrees
of freedom in the irrep $R$.
The barred operator contains spin
components of the fermion field
$\overline{q}(x)$, which lies in the anti-fundamental
representation of colour $SU(3)$, and the unbarred operator
contains the spin components of $q(x)$.
$\mathcal{W}(t+\tau;t,\vec{x})$ is a
Wilson line from $t$ to $t+\tau$ at the
spatial site $\vec{x}$.
In Eq.(\ref{eqn:corr_matrix}) only the contribution to the correlator involving
the lower quark field components is given explicitly, and the initials
T.R. denote the piece
involving the upper quark field components, obtained by  
time reversal.
In terms of the interpolating operators appearing in 
Eq.(\ref{eqn:corr_matrix}), the operator which optimally couples 
to the $n^{th}$ excited state is given by
\begin{eqnarray}
\phi_{n}^{(R)}(t,\vec{x}) = \sum_{\alpha} v_{n \alpha}
\mathcal{O}^{(R)}_{\alpha}(t, \vec{x}).
\end{eqnarray}
The real-valued coefficients $v_{n \alpha}$ form the
elements of a vector $\mathbf{v}_{n}$ which
is obtained by solving the generalised eigenvalue equation
\begin{eqnarray}
C(t_{D}) \mathbf{v}_{n} =
\lambda_{n}(t_{D},t_{0}) C(t_{0}) \mathbf{v}_{n},
\label{eqn:gen_eigenval}
\end{eqnarray}
where $t_{0}$ is a fixed initial time and
$t_{D}$ is a later reference time. The
eigenvalues in Eq.(\ref{eqn:gen_eigenval}) satisfy 
$\lim_{t \rightarrow \infty} \lambda_{n}(t,t_{0}) = e^{-E_{n} (t -
t_{0})} \left[ 1 + \mathcal{O} \left( e^{- \delta_{n}(t - t_{0})}
\right) \right]$, where $E_{n}$ is the energy of the $n^{th}$ 
excited state and $\delta_{n}$ is the absolute value of the energy 
difference to the nearest state.

The excited state energies can be determined by fitting to
optimised correlation matrices
\begin{eqnarray}
\widetilde{C}_{n m}(\tau) &\propto&
\sum_{t, \vec{x}}
\langle \Omega |
\overline{\phi}^{(R)}_{n} ( t + \tau, \vec{x})
\mathcal{W}(t+\tau,t;\vec{x})
\phi_{m}^{(R)}(t, \vec{x}) | \Omega \rangle \nonumber  \\ 
& & + \rm{T.R.} , \nonumber \\
&\equiv& \mathbf{v}_{n} C(\tau) \mathbf{v}_{m}.
\label{eqn:op_corr_matrix}
\end{eqnarray}

To construct the initial correlation matrices, $C(\tau)$, we require a set of
operators which transform according to the irrep $R$.
However, these operators cannot lie in a 
single instance of $R$ because the correlation between 
different rows in a single instance of an irrep 
is zero, rendering the variational method invalid.
To obtain correlation matrices we apply a number of levels
of Jacobi smearing to the quark field components used in the interpolating
operators which generates several instances of the irrep. For
each row in $R$, cross correlations of
operators appearing in different instances of the irrep are computed.
This results in a correlation matrix for each row
in the irrep. These matrices are identical up to statistical errors and
we average over the rows of the irrep to obtain a final
correlation matrix.
The variational method (Eq.(\ref{eqn:gen_eigenval}), Eq.(\ref{eqn:op_corr_matrix})) is then applied to
this averaged correlation matrix to determine the energy of the ground state
as well as higher-lying states in $R$.

%% file: all-to-all.tex
\subsection{All-to-all propagators}
\label{all2all}
The use of an efficient estimate for all elements of the lattice
Dirac propagator lies at heart of this study.
Although the move from point-to-all fermion propagators to all-to-all
propagators undoubtedly introduces an additional
computational overhead into lattice simulations, there exist
very many applications in lattice field theory where the benefit of
using all-to-all propagators far outweighs this additional cost.
The work described here involves two areas where all-to-all
techniques  are particularly useful. Firstly, the reduction in noise
per configuration achieved by averaging the correlation matrices over
the lattice volume is essential in studies of static-light systems
on rather expensive dynamical background configurations.
Secondly, all-to-all propagators are of great advantage in excited
state spectroscopy. 
Because all elements of the lattice propagator are known, 
the construction of spatially-extended  interpolating operators
and the smearing of the quark fields require no additional inversions
of the fermion matrix. 
The utility of all-to-all propagators in 
constructing operators which optimally couple 
to orbital and radial excitations cannot be overstated. 

To estimate all elements of the Dirac propagator 
we employ the hybrid method introduced in ref.~\cite{Foley:2005ac}. 
In this approach, the contribution of the low-lying 
eigenmodes of the Dirac operator to the fermion 
propagator is evaluated exactly, since these modes 
are thought to dominate the long-range physics of interest; 
the contribution of the higher eigenmodes is then 
estimated stochastically.
In terms of the hermitian fermion matrix $Q = \gamma_{5} M$, 
the propagator 
is simply 
\begin{eqnarray}
M^{-1} = Q^{-1} \gamma_{5}. 
\end{eqnarray}
To proceed, $Q^{-1}$ is written as a spectral sum which is divided 
into two pieces
\begin{eqnarray}
Q^{-1} = \overline{Q}_{0} + \overline{Q}_{1} ,
\end{eqnarray}
with
\begin{eqnarray}
\overline{Q}_{0} = \sum_{i=1}^{N_{ev}} \frac{1} {\lambda_{i}} v^{(i)} \otimes v^{(i)\dagger}, 
\hspace{2mm} \overline{Q}_{1} = \sum_{i = N_{ev}+1}^{N} \frac{1}
{\lambda_{i}} v^{(i)} \otimes v^{(i)\dagger}  ,
\end{eqnarray}
where $Q v^{(i)} = \lambda_{i} v^{(i)}$ and $N$ is the rank of the 
fermion matrix. Therefore, according to our method,
$\overline{Q}_{0}$, which contains the first $N_{ev}$ eigenvectors, is 
computed exactly. $\overline{Q}_{1}$ is the contribution of the 
higher-lying modes to the propagator,  which is estimated stochastically. 
Using the  projection operator, 
$P_{1} = 1 - \sum_{i=1}^{N_{ev}} v^{(i)} \otimes v^{(i) \dagger}$, 
this contribution can be written in a more useful form 
$\overline{Q}_{1} = Q^{-1} P_{1}$.  

In a naive stochastic estimate, one generates a set of $N_{r}$ random noise 
sources $\{\eta_{[1]}....\eta_{[N_{r}]}\}$ with the structure of quark fields 
and the property that 
\begin{eqnarray}
\langle \langle \eta_{[r]}(x) \otimes \eta_{[r]}(y)^{\dagger} \rangle
\rangle \approx \delta_{x y},
\end{eqnarray}
where $x$, $y$ denote any set of quark field indices and 
$\langle \langle ... \rangle \rangle$ denotes an average over the set of noise vectors. 
An estimate for $\overline{Q}_{1}$ is then given by 
\begin{eqnarray}
\overline{Q}_{1}(x,y) \approx \langle \langle \psi_{[r]}(x) \otimes \eta_{[r]}^{\dagger} \rangle \rangle ,
\label{eqn:naive_est}
\end{eqnarray}
where the set $\{ \psi_{[r]} \}$ is obtained by solving
$\psi_{[r]} = Q^{-1} P_{1} \eta_{[r]}$ for each source vector. 
Unfortunately, the variance in this estimate can be quite large, 
and this error decreases as $1 \sqrt{N_{r}}$ so that Eq.(\ref{eqn:naive_est}) becomes exact only 
in the limit $N_{r} \rightarrow \infty$. 
However, the stochastic estimate can be improved upon substantially through 
a careful partitioning or `dilution' of the noise sources.   
The idea here is to break each original noise source 
into disjoint pieces according to some set of quark field 
indices. For example, in time dilution, which appears to play a particularly 
important role in variance reduction, each noise vector is partitioned 
as
\begin{eqnarray}
\eta_{[r]}(\vec{x}, t) = \sum_{i=0}^{N_{t}-1} \eta^{(i)}_{[r]}(\vec{x},t),
\end{eqnarray}
where $\eta^{(i)}(\vec{x},t) = 0$ for $ i \neq t$. 

One can dilute in any combination 
of quark field indices, generating $N_{d}$ diluted noise vectors 
from each original source. Inverting the fermion matrix on these 
diluted sources yields an unbiased estimate for $\overline{Q}_{1}$ 
from a single original noise source 
\begin{eqnarray}
\sum_{i=0}^{N_{d} - 1} \psi^{(i)}_{[r]}(\vec{x},t) \otimes \eta^{(i) \dagger}_{[r]}(\vec{x}_{0}, t_{0}).
\end{eqnarray} 
In this study, we use $Z_4$ noise and 
each component of the noise sources has modulus 1. 
For such noise, the limit of full dilution
corresponds to an exact evaluation of $\overline{Q}_{1}$. 
This is in stark contrast to the naive estimate  where an infinite number 
of matrix inversions is required to evaluate the propagator exactly.
  
Averaging each diluted estimate over the distribution of noise 
sources means that a total of $N_{inv} = N_{d} \times N_{r}$ 
matrix inversions are used in the estimator of $\overline{Q}_{1}$. 
There is considerable freedom in how we partition 
the noise vectors and the optimal dilution path is that dilution which yields satisfactory 
results while minimising $N_{inv}$.   

Although the number of eigenvectors, the number of source vectors and the 
dilution level required depend very sensitively on the details of the 
measurement, in our studies of static-light systems 
we have found that computing the contribution of a small number 
of the low-lying eigenmodes exactly, and using just one or two 
noise sources with a moderate level of dilution yields excellent 
results. The precise details of the fermion propagator estimates 
used in this study are given in Section~\ref{details}.

%% file: latticeactions.tex
\subsection{Lattice actions}
\label{actions}
The lattice gauge and Dirac actions used in this study are
described in detail in refs.~\cite{Foley:2004jf,Morningstar:1999dh}. 
They are formulated specifically
for anisotropic lattices. The gauge action is 
Symanzik and tadpole improved with leading discretisation errors of 
${\cal O}(a_s^4, a_t^{2}, \alpha_sa_s^{2})$, where $a_s$ is the lattice
spacing in the spatial directions and $a_t$ is the lattice spacing in
the temporal direction. 
The fermion action is Wilson-like in the temporal direction 
but the lattice Lagrangian includes an operator of 
dimension seven to lift the spatial doublers. This 
action has leading discretisation errors
of ${\cal O}(a_s^{3}, a_t, \alpha_s a_s)$.
In addition,  
the spatial links appearing in 
the fermion action are stout-smeared to reduce radiative corrections. 

The anisotropic lattice breaks hypercubic symmetry and 
introduces two new parameters: the bare quark and 
gluon anisotropies, denoted $\xi_{q}^{0}$ and $\xi_{g}^{0}$, which 
must be tuned so that the measured anisotropy $\xi = a_{s}/a_{t}$ takes 
a target value, which must be independent of the physical  
probe used to perform the measurement. In dynamical QCD the tuning 
procedure is quite complicated but is now well-understood 
and is described in ref.~\cite{Morrin:2006tf}.

The static quark is simulated with the Eichten-Hill action~\cite{Eichten:1989kb} and the
static quark propagator is then the product of unsmeared temporal
links. 

%% file: simulationdetails.tex
\subsection{Simulation details}
\label{details} 
The study was performed at a single lattice spacing with $N_f=2$. Some
details of the simulation, for both volumes, are in
Table~\ref{tab:latticedetails}. 
\begin{table}[h]
\begin{center}
\begin{tabular}{ccccccc}
\hline
Volume         & $\beta$ & $a_tm_q$ & $a_s$ & $\xi_q^0$ & $\xi_g^0$ 
	                                                     & Configs
	                                                     \\
\hline
$8^3\times 80$ & 1.508 & -0.057 & $\sim$0.17fm& 7.43 & 8.42 & 232 \\ 
$12^3\times 80$& 1.508 & -0.057 & $\sim$0.17fm& 7.43 & 8.42 & 245 \\ 
\hline
\end{tabular}
\caption{A table of simulation parameters for the two volumes used in
this study. }
\label{tab:latticedetails}
\end{center}
\end{table}
The lattice spacing is the same on both volumes and the light and sea
quarks are degenerate in the simulation. 
The determination of the bare quark anisotropy, $\xi_{q}^{0} = 7.43$ and the
bare gluon anisotropy, $\xi_g^0=8.42$ is described in
ref.~\cite{Morrin:2006tf}. These values were chosen to obtain a physical 
anisotropy $\xi = 6$. Measurements of the anisotropy from the light
pseudoscalar dispersion relation on the small 
and large volumes yield values of $5.94(6)$ and $5.82(5)$ respectively, 
which are consistent with the analysis of ref.~\cite{Morrin:2006tf}.
For this parameter set, the ratio $m_{\pi}/m_{\rho} = 0.54$. Our study
is therefore most applicable to the spectrum of $B_{s}$ mesons. 
The temporal lattice spacing, determined from the spin-averaged 1P-1S
splitting in charmonium, on the same ensemble~\cite{Juge:2006fm}, 
was found to be
$a^{-1}_{\rm 1P-1S} = 7.04(3)$GeV, which implies a value for the
spatial lattice spacing, $a_{s}$ of approximately  $0.17~\rm{fm}$. 
Simulations were performed on $8^{3} \times 80$ and $12^{3} \times 80$ 
lattices. The use of two spatial volumes allowed us to 
identify possible finite-size effects and probable multiparticle 
excitations in our results.  

The gauge configurations were generated using the standard Hybrid 
Monte Carlo (HMC) algorithm. A more detailed description of the 
particular improvements to the HMC algorithm, specifically for 
anisotropic lattices is in reference~\cite{Morrin:2006tf}. 
The generation of the ensemble of 232 gauge configurations on 
the small lattice required 
approximately 5000 CPU hours while the 245 configurations on the 
larger lattice were generated in approximately 15000 CPU hours. 

On each volume, we computed the exact contribution of the 50 lowest-lying 
eigenmodes to the light quark propagator using the ARPACK implementation 
of the Lanczos algorithm to determine the
eigenvectors.  
The choice $N_{ev}=50$ was based on earlier analyses carried out on the small volume. 
For similar run parameters, it was found that computing the contribution 
of a few low-lying eigenmodes to the fermion 
propagator exactly did result in a significant 
improvement in signal resolution compared to 
a purely stochastic estimate for the propagator.  
However, the gain in precision achieved by including more 
eigenmodes quickly decreases so that, for our purposes, it was not worth 
considering $N_{ev}$ greater than 50. 
To evaluate the contribution of the higher modes 
to the light fermion propagator on the small 
volume, time and colour dilution was applied 
to two independent noise sources. On the larger 
volume, a single noise source was used, which was  
also diluted in time and colour indices. 

Once the light fermion propagators had been computed, 
the evaluation of the correlation matrices was relatively straight-forward.
Regarding the use of stout-links in the interpolating operators,
we found that 
satisfactory results were obtained with 30 iterations 
of the smearing algorithm using a weighting factor 
of $\rho = 0.025$.  
In order to construct correlator matrices, 8 levels of Jacobi 
smearing were applied to the light quark fields. 
The quark field after $j$ applications of the smearing algorithm is given by 
\begin{eqnarray}
q^{(j)}(x) = \left(  1 + \kappa \sum_{i = 1}^{3} D_{i}^{2} \right)^{n_{\kappa}} q^{(j-1)}(x),
\end{eqnarray} 
where $q^{(0)}(x)$ is the original unsmeared quark field. 
On both lattices, $\kappa$ was chosen to be $0.1$ and $n_{\kappa}$ 
was fixed at 12. 
For the spatially-extended interpolating operators, 
we varied the lengths of the gauge-covariant paths, testing 
offsets which extended over a single lattice spacing and 
two lattice spacings in a given direction. We found that the 
single-link operators coupled optimally to the ground states 
in each lattice irrep and these are the operators which were used 
in our final measurements.

%% file: results.tex
\section{Results}
\label{results}
\begin{figure}[h]
  \centering
  \includegraphics*[width=7cm]{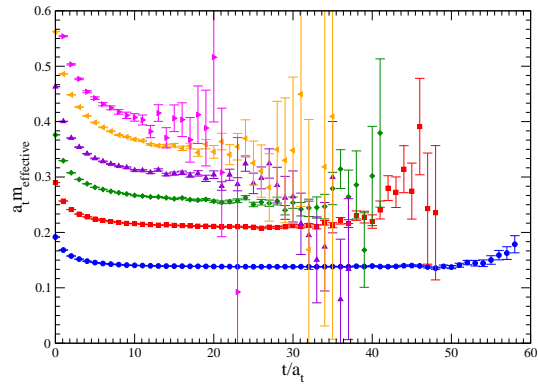}
  \caption{The effective masses in the $G_{1u}$ irrep, containing
   the static-light S wave, determined from the optimised
  correlation matrix. The ground state and 5 excitations are shown. }
  \label{fig:swave-meff}
\end{figure}

\begin{figure}[h]
  \centering
  \includegraphics*[width=7cm]{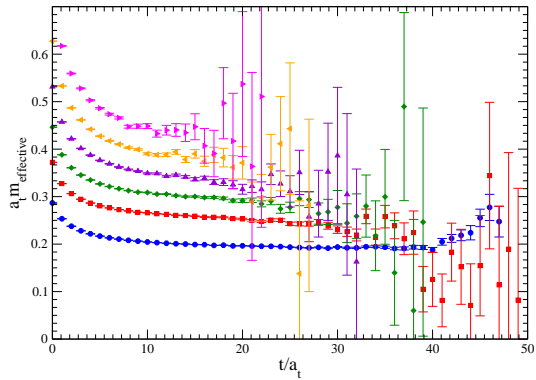}
  \caption{The effective masses in the $G_{1g}$ irrep, corresponding
  to the $\frac{1}{2}^+$ P wave. The ground state and 5 excitations
  are shown.}
  \label{fig:g1wave-meff}
\end{figure}

\begin{figure}[h]
  \centering
  \includegraphics*[width=7cm]{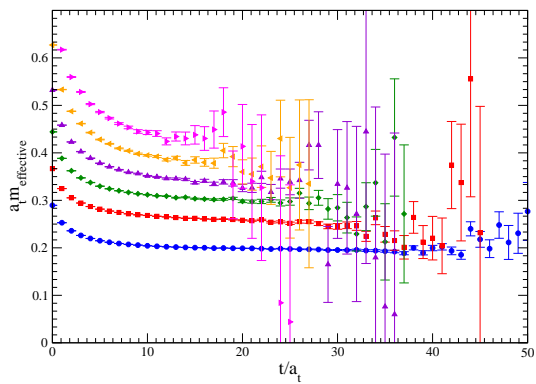}
  \caption{The effective masses in the $H_g$ irrep, associated with
  the $\frac{3}{2}^+$ P wave. The ground state and five excitations
  are shown.}
  \label{fig:hpwave-meff}
\end{figure}

\begin{figure}[h]
  \centering
  \includegraphics*[width=7cm]{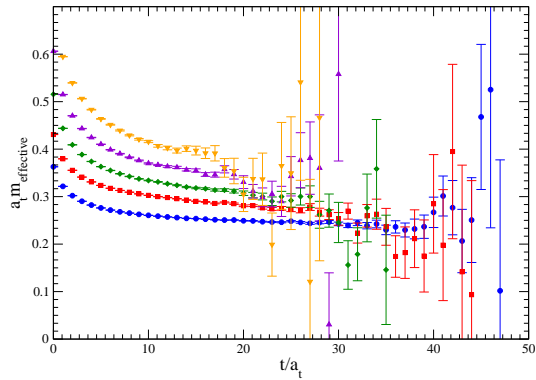}
  \caption{The effective masses in the $H_u$ irrep, which contains 
  the $\frac{3} {2}^{-}$ and $\frac{5} {2}^{-}$ D-wave states. The ground state and four excitations are shown.}
  \label{fig:hdwave-meff}
\end{figure}

\begin{figure}[h]
  \centering
  \includegraphics*[width=7cm]{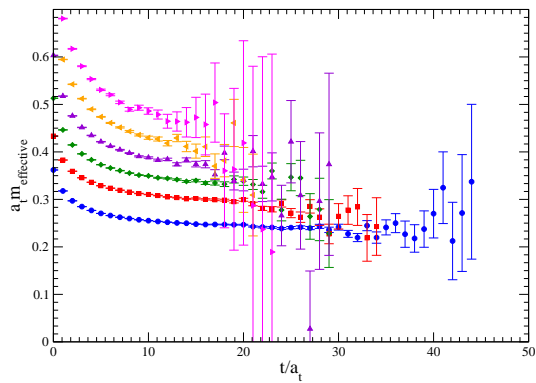}
  \caption{The effective masses in the $G_{2u}$ irrep, whose lowest
  lying state is a D wave. The ground state and five excitations are shown.}
  \label{fig:g2wave-meff}
\end{figure}

In Figures~\ref{fig:swave-meff} to \ref{fig:g2wave-meff} we plot the
effective masses, given by 
\begin{equation}
m^{(n)}_{\rm effective} = 
\ln\left[ \widetilde{C}_{nn}(t)/\widetilde{C}_{nn}(t+1)\right] ,
\label{eqn:effmass}
\end{equation}
for the correlation matrices in each of the five representations
considered. These results on the larger ($12^3\times 80$) lattice show
the high quality of the data. Convincing plateaux, with small
statistical errors, are seen for as many as five or six states in each
irrep. While the physical significance of the higher-lying levels
cannot be pinned down without further study the very precise data give
us confidence in our determination of the lower-lying energy states. 
\subsection{Fitting Procedures}
The vectors, $\mathbf{v}_{n}$, used to determine the optimised 
correlation matrices were obtained by solving the 
eigenvalue problem in Eq.(\ref{eqn:gen_eigenval})
on time slices $t_{0}=2$ and $t_{D}=10$, although 
we tested other time slices to check for 
consistency. The resulting correlation matrices are then 
guaranteed to be diagonal only on time slice 2 and time slice 10.
However, for high-quality data we expect the 
off-diagonal entries of the optimised correlation matrices to be very
small over some time interval. 
In this interval, estimates for the excited state energies 
can be obtained by fitting to the diagonal entries in the correlation
matrices. The effective masses in Figures~\ref{fig:swave-meff} to 
\ref{fig:g2wave-meff}, which are defined in terms of the diagonal
entries of the correlation matrix, demonstrate that this assumption is
reasonable. 

However, the most reliable estimates for the low-lying energies 
are found by including off-diagonal matrix entries in the fit. 
To implement this, submatrices involving operators which project onto
a smaller number, M, of the lowest-lying states are extracted from the
full $8\times 8$ optimised matrices.
The energies of these states $E_{n}$ are determined by fitting to the
entries of this submatrix $\widetilde{C}_{PQ}(t)$ using the ansatz 
\begin{eqnarray}
\widetilde{C}_{P Q}(t) = \sum_{n=0}^{M-1}Z_n^PZ_n^Q e^{-E_{n} t} .
\end{eqnarray}
On the small volume, we fit to a $3\times3$ submatrix 
in the $G_{1u}$ irrep, and to $2 \times 2$ matrices 
in the other irreps. The enhanced signals on the larger 
volume meant that it was possible to fit to a 
$4 \times 4$ matrix in the $G_{1u}$ irrep and $3 \times 3$ correlation 
matrices in the other representations. All the fits use a
correlated $\chi^2$-minimisation algorithm and statistical errors are determined
from 1000 bootstrap samples. 

\begin{figure}[h]
  \centering
  \includegraphics*[width=7cm]{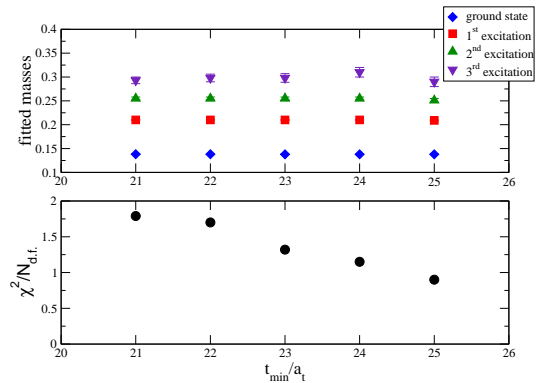}
  \caption{A sliding window plot for the $G_{1u}$ irrep. The maximum
   time slice for these fitted masses is $t_{\rm max}=30$. The fitted
   masses in the ground and first three excited states are stable,
   with good $\chi^2/{\rm N_{\rm d.f.}}$, when
   $t_{\rm min}$ is varied in fits to the full correlation matrix
   using four exponentials. Although the range of stable values of
   $t_{\rm min}$ is smaller than seen in
   Figure~\ref{fig:sliding-windowHd} the number of states determined
   is larger. }
  \label{fig:sliding-windowS}
\end{figure}

\begin{figure}[h]
  \centering
  \includegraphics*[width=7cm]{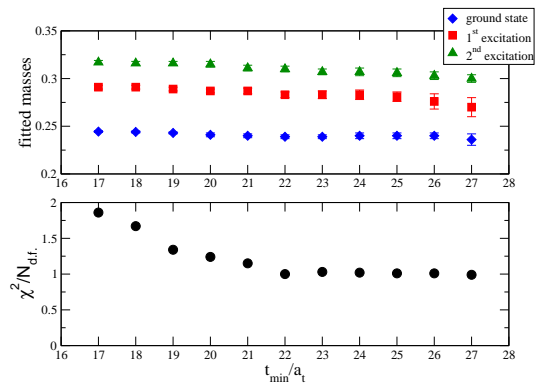}
  \caption{A sliding window plot for the $H_u$ irrep. Here $t_{\rm
  max}=35$ and the ground and first excited states are seen to be
  stable, with good $\chi^2/{\rm N_{\rm d.f.}}$, with respect to changes in
  $t_{\rm min}$.}
  \label{fig:sliding-windowHd}
\end{figure}

Since the fitting routine allows for the fact that the operators  
${\mathbf v}_n$ may not be orthogonal, it permits   
us to fit to time ranges much greater than the optimisation 
time $t_{D}$. The optimal fit ranges were determined 
from a sliding window analysis where $t_{max}$ was fixed 
at an appropriate value and $t_{min}$ was varied; 
the stability of the fitted values and the goodness 
of fit for varying $t_{min}$ were then examined.    
Sample sliding window plots, for the S wave irrep and the $H_{u}$ irrep,
are shown in Figures~\ref{fig:sliding-windowS} and \ref{fig:sliding-windowHd}. 
The fitted masses are seen to be consistent over a range of values of
$t_{\rm min}$ with a good
$\chi^2/{\rm N_{\rm d.f}}$. Similar results were found for the five irreps
considered. The final fit ranges and the best-fit energies 
for each of the lattice irreps are given in Table~\ref{tab:bestfits}.
\begin{table}[h]
\begin{center}
\begin{tabular}{cccccc}
\hline
Irrep & ($t_{\rm min}$, $t_{\rm max}$) & \multicolumn{3}{c}{Channels} 
           & $\chi^2/{\rm N_{\rm d.f.}}$  \\
&          & ground & $1^{st}$ & $2^{nd}$   &          \\
\hline
$G_{1u}$ & (24,30)& 0.1379(2)  &0.210(1)  &  0.255(3)  &  1.15  \\
$G_{1g}$ & (25,30)& 0.1906(7)  & 0.246(4)  &  0.291(6) &  1.37  \\
$H_{g}$ & (25,30) & 0.1956(7)  & 0.257(2)  & &  1.02  \\
$H_{u}$ &(22,35) & 0.239{(2)\phantom{1}}  & 0.283(3)  & &  1.00  \\
$G_{2u}$ &(24,30) & 0.230{(4)\phantom{1}} & 0.283(8)  & & 1.35  \\
\hline
\end{tabular}
\caption{The best-fit energies of the ground states and higher
  excitations of each of the five irreps considered. The energies 
  are given in units of $a_{t}^{-1}$. The fit ranges
  used and the $\chi^{2}/{\rm N_{\rm d.f.}}$ values are also given. These results
  are obtained on the larger of our two volumes, $12^3\times 80$. The
  errors are statistical only.}
\label{tab:bestfits}
\end{center}
\end{table}
In each of these fits, we disregard the result for the highest energy
included in the fit since we expect that result to suffer from contamination 
from higher levels. As an example, for the S-wave ($G_{1u}$) irrep we
believe the ground and first two excited states are convincingly
determined in our fit procedure. The signal for the remaining
higher-lying excitations is still remarkably good, as shown in the
effective mass plots and we estimate the energies of these states
using single and double exponential fits to the relevant correlation
matrices, 
\begin{eqnarray}
\widetilde{C}_{nn}(t) &=& A_ne^{-E_nt} ,\\
\widetilde{C}_{nn}(t) &=& A_ne^{-E_nt}\left(1+B_ne^{-\Delta_nt}\right) .
\end{eqnarray}
All the fits  were done using a correlated $\chi^{2}$-minimisation algorithm.
The statistical errors on the fits were again estimated from 1000 
bootstrap samples. 
Using the double exponential form it was possible to fit to relatively
early time ranges, while the single exponential was applicable to later time intervals. 
We have checked that fitting these ans\"atze over different time
intervals yields consistent results for all the lattice irreps. In
addition we have applied this fitting procedure to the lower-lying
energies and found that the resulting fitted energies agree within
errors with the energies determined using the full correlation matrix,
as described above. 

We have also compared the effective masses defined in Eq.(\ref{eqn:effmass}) 
to effective masses obtained by diagonalising 
the correlation matrices on each time slice, as proposed in
ref.~\cite{Luscher:1990ck} and found that both approaches agree over the fit intervals.

\subsection{The spectrum}
\begin{figure}[h]
  \centering
  \includegraphics*[width=8cm]{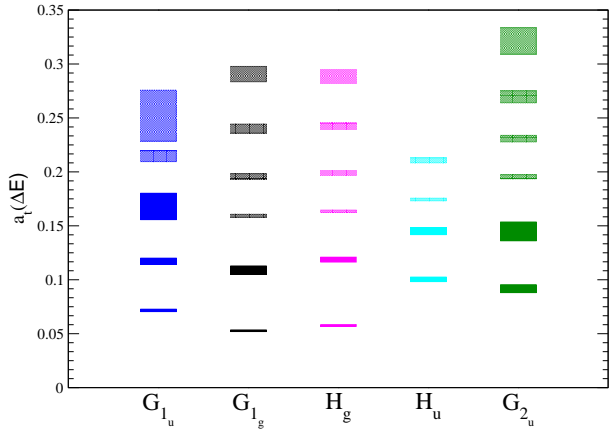}
  \caption{The splitting between the ground state S wave (ground state
  in the $G_{1u}$ irrep) and the other states determined in this
  study. We show the splitting between states determined from fits to
  the full correlation matrix as solid bands. Such fits ensure that
  even small off-diagonal elements of the correlation matrix, which
  represent leaking between the levels, are taken into account. The
  shaded bands are those splittings determined, for the higher-lying
  states, using fits to the diagonal elements of the correlation
  matrix only. As discussed in the text the data indicate that this is
  a reasonable procedure, nevertheless we present these states as less
  conclusively determined. The possible inversion of ordering seen in
  the D-wave orbital is discussed further in
  Section~\ref{discussion} of the text.}
  \label{fig:spectrum}
\end{figure}

In the static limit, hadron energies receive 
an unphysical contribution from the static quark self-energy. 
However, this contribution cancels in energy differences 
which are therefore of direct physical significance. 
The energy differences between the ground state in $G_{1 u}$, which is the S wave,   
and higher states are shown in Figure~\ref{fig:spectrum}. 
The solid bars are those states determined by a fit to the full
correlation matrix. In this case we are satisfied that these numbers
correspond to the eigenstates of the lattice Hamiltonian. The dashed
boxes denote energy levels obtained by fitting to the diagonal
elements of the correlation matrix as described earlier. Although the 
data indicate that the dominance of the diagonal entries is a
reasonable assumption we nevertheless present these results as less
conclusive. Table~\ref{tab:phys-splittings} lists the splittings shown
by the solid bands in Figure~\ref{fig:spectrum} in physical units. 
\begin{table}[h]
\begin{center}
\begin{tabular}{cc}
\hline
 channel & $\triangle E$ (GeV) \\
\hline
$G_{1u}$, 1st excitation & 0.504(8) \\
$G_{1u}$, 2nd excitation & 0.82{(2)\phantom{1}} \\
\hline
$G_{1g}$, ground state & 0.371(6)\\
$G_{1g}$, 1st excitation & 0.76{(3)\phantom{1}} \\
\hline
$H_{g}$, ground state & 0.405(6)\\
$H_{g}$, 1st excitation & 0.84{(2)\phantom{1}}\\
\hline
$H_u$, ground state & 0.706(2)\\
$H_{u}$, 1st excitation & 1.02{(2)\phantom{1}}\\
\hline 
$G_{2u}$, ground state & 0.65{(3)\phantom{1}}\\
$G_{2u}$, 1st excitation & 1.02{(6)\phantom{1}}\\
\hline
\end{tabular}
\caption{The splitting between the ground state in the $G_{1u}$ (S-wave)
  irrep and the other states determined in this analysis by fits to
  the full optimised correlation matrices. The lattice spacing was
  determined from the spin-averaged 1P-1S splitting in charmonium, on
  the same ensemble.  
}
\label{tab:phys-splittings}
\end{center}
\end{table}
The lattice spacing has been set from the spin-averaged 1P-1S
splitting in charmonium determined on the same configuration set and
described in ref.~\cite{Juge:2006fm}.

%% file: finitevolume.tex
\subsection{Finite volume effects}
The first indication that finite volume  
effects might be significant comes 
from the slight discrepancy between the quark anisotropy 
on the small and large volumes. 
The small lattice used in this study has 
a spatial volume of about $(1.35~\rm{fm})^{3}$.  
The spatial volume of the large lattice is 
$(2.03~\rm{fm})^{3}$. 
Comparing the lowest energy levels 
in each of the irreps across both volumes, 
we find small but significant shifts in the energies 
of the $G_{1u}$, $G_{1g}$ and $H_{g}$ states. 
The lowest energy levels in the $H_{u}$ and 
$G_{2 u}$ representations remain constant 
within statistical errors. 
The $G_{1u}$ energy level decreases slightly 
on the larger volume.   
The energies of the $G_{1 g}$ and $H_{g}$ states  
show the strongest volume dependence. 
The energies of these states increase on the large volume 
and the splitting between them decreases. This is illustrated in
Table~\ref{tab:finitevol} which shows the fitted masses, for the
lowest-lying state in each irrep on two volumes. 
\begin{table}[h]
\begin{center}
\begin{tabular}{ccc}
\hline
Irrep & \multicolumn{2}{c}{Energies}  \\
      &       $8^3\times 80$ & $12^3\times 80$  \\
\hline
$G_{1u}$& 0.1411(5) & 0.1379(2) \\
$G_{1g}$& 0.174{(1)\phantom{1}}  & 0.1906(7) \\
$H_g  $ & 0.183{(2)\phantom{1}}  & 0.1956(7)\\
$H_u  $ & 0.233{(3)\phantom{1}}  & 0.239{(2)\phantom{1}}\\
$G_{2u}$& 0.224{(3)\phantom{1}}  & 0.230{(4)\phantom{1}} \\
\hline
\end{tabular}
\caption{A comparison of the fitted energies of the lowest-lying states in
each irrep on two volumes. These values are determined in similar fits
to the full correlation matrix. The largest effect is seen in the P
wave states. The energies of these states increase on the larger
volume while the splittings between them decrease. }
\label{tab:finitevol}
\end{center}
\end{table}
These states are expected to correspond to P wave 
excitations of the static-light meson.  
Although a noticeable volume dependence in energies can 
be a signal for scattering states involving particles 
with non-zero spatial momentum, one expects the energies 
of such states to decrease on a larger spatial volume. 
Furthermore, unless the ground state energies in  $H_{u}$ and 
$G_{2 u}$ also correspond to multiparticle states, such an assignment 
is highly unlikely. 
The possibility that the observed energy levels correspond 
to two-particle scattering states will be discussed 
in further detail in Section.~\ref{multiparticle}

%% file: massandspacing.tex
\subsection{Quark mass and lattice spacing effects}
\label{massandspacing}
The results presented here have been obtained on gauge 
configurations generated with a probability density  
which includes the effects of two degenerate sea quark flavours. 
The mass of the sea quarks was close to the strange quark mass 
and the same value was used for the mass of the light valence quark. 
Therefore, we have simulated a well-defined unitary  
field theory and our results are free of quenched
pathologies. However, we have not explored
the inclusion of additional flavours or the use 
of lighter quark masses. 
Recent developments in algorithms indicate that these improvements  
can be implemented with moderate computing resources using 
the same actions in the near future~\cite{DelDebbio:2005qa}. 

The largest uncertainty in our results is due to 
the fact that we have carried out simulations at just a 
single lattice spacing. Therefore, although we are confident 
that the energies we have computed correspond to 
eigenstates of the lattice Hamiltonian, we are unable to 
extrapolate these results to the continuum limit.   
Moreover, the spatial lattice spacing used in the simulations was 
quite coarse,
$a_{s} \sim 0.17~\rm{fm}$, and cutoff effects 
in our results may be large although we hope that the 
use of improved actions keeps these errors under control.

%% file: multiparticle.tex
\section{Multiparticle states}
\label{multiparticle}
It is possible that a number 
of the energy levels we have computed are not the 
energies of bound-state resonances but correspond 
to multiparticle states.  
The pion mass measured on this configuration set 
is $a_{t} m_{\pi} \sim 0.05579$ and one can expect two-particle states 
consisting of a static-light meson and a pion in particular to 
contribute to the energy regime shown in Figure~\ref{fig:spectrum}. 
Since we have obtained results on two spatial volumes, 
it should be possible to identify multiparticle resonances by comparing 
the elements of the optimisation vectors $\mathbf{v}_{n}$, which 
are computed in the variational method, on both volumes.  
The normalisation condition for these vectors is $\mathbf{v}^{\rm{T}}_{n} C(t_{0}) \mathbf{v}_{n} = 1$. 
A vector projecting onto a bound state is, to a good approximation, 
volume independent. On the other hand, a vector which isolates a multiparticle 
state shows a strong, well-defined volume dependence~\cite{Mathur:2004jr}. 
This dependence comes from the fact that, for example, a state 
consisting of two weakly-interacting particles in a finite 
spatial volume has a spectral weight which scales with the inverse 
volume. The ratio of the volumes used in this study was 3.375, implying that 
multiparticle excitations should be readily identifiable.  
However, studies of possible pentaquark candidates have  
shown that considerable care must be taken when trying to separate 
single-particle energies from the energies of scattering states on the basis 
of the volume dependence of 
spectral weights~\cite{Liu:2005yc,Alexandrou:2005ek,Alexandrou:2005gc}. 
In particular, when attempting to isolate the spectral weight of a single state 
one needs to check for contamination from higher states.
We therefore consider the volume dependence of the 
eigenvectors in conjunction with fit values for the overlap of the optimised 
interpolating operator onto the state under consideration. 
We examine the eigenvectors which project onto the three lowest 
states in the $G_{1u}$ irrep and the two lowest states 
in each of the other irreps, for which we have reliable energy measurements. 
However, we find no evidence for multiparticle contributions 
to the energy spectrum shown in Fig.~\ref{fig:spectrum}, although such states 
undoubtedly exist. 
We therefore conclude that the interpolating operators 
used have negligible overlap with low-lying multiparticle resonances.   
Perhaps this is not so surprising, by construction one expects 
the interpolating operators to couple strongly to just a few low-lying 
single-particle states.  
In fact, given that the operators used in the 
variational bases differed only in the smearing of the quark fields, 
it is rather remarkable that the variational method appears to yield 
good results for so many single-particle resonances. 

Ultimately, the issue of multiparticle states in the static-light 
spectrum will be resolved by using much larger operator bases 
incorporating interpolating operators specifically 
chosen to project onto multiparticle states. 
However, using the data available to us it is possible to make some
simple predictions for the multiparticle spectrum.
If we assume that the interaction between particles in a scattering state is very weak, 
we can determine approximate values for the threshold energies for multiparticle states in each of the 
irreps. 
For a large enough spatial volume  
the lowest-energy two-particle states consist of the 
ground-state S-wave static-light meson and a pion which may have non-zero momentum. 
We use periodic boundary conditions in the spatial directions 
so the components of the pion momentum $ \vec{p} $ are quantised in units of 
$2\pi/a_{s}L$. 
A pion at rest transforms according to the $A_{1 u}$ irrep of 
$O_{h}$. However, by applying the elements of $O_{h}$ to a
pion field with non-zero momentum, one can generate more complicated 
single-valued representations of the group. Table~\ref{tab:pion} 
gives the decomposition of pion representations into their constituent 
irreps for the smallest values of $|\vec{p}|$ allowed by the lattice.
\begin{table}[h!]
\centering
\begin{tabular}{c|l}
\hline
$\vec{p}$ & Irreducible content \\
\hline
$ (0,0,0) $ & $A_{1 u}$ \\
$ (1,0,0) $ & $A_{1u} \oplus E_{u} \oplus T_{1g} $ \\
$ (1,1,0) $ & $A_{1u} \oplus E_{u} \oplus T_{1g}  \oplus T_{2g} \oplus T_{2u} $ \\
\hline
\end{tabular}
\caption{An example of the decomposition of pion representations of $O_{h}$, labeled by the pion momentum 
in units of $2\pi/a_{s}L$, into their constituent irreps.}
\label{tab:pion}
\end{table}
Representations for the two-particle scattering states of interest are given by the 
direct product of the pion representations with $G_{1 u}$.  
We then decompose these representations in terms of the double-valued 
irreps of $O_{h}$. By considering different pion momenta one 
can deduce the two-particle scattering content of each of the irreps.   
This procedure is in complete analogy to the determination of the angular momentum 
content of the irreps described in section~\ref{grouptheory}. 
The energy of a scattering state is approximated by the 
sum of the energies of the decoupled states.
\begin{eqnarray}
E_{\rm{2P}} \approx E_{S.L.} + \sqrt{ m_{\ell}^{2} + p^{2}},
\label{eqn:threshold}
\end{eqnarray}
where $m_{\ell}$ is the rest mass of the light meson appearing in the scattering state. 
The threshold for multiparticle states in a particular irrep 
is the minimum such energy appearing in that representation. 

\begin{figure}[h]
   \centering 
   \includegraphics*[width=7cm]{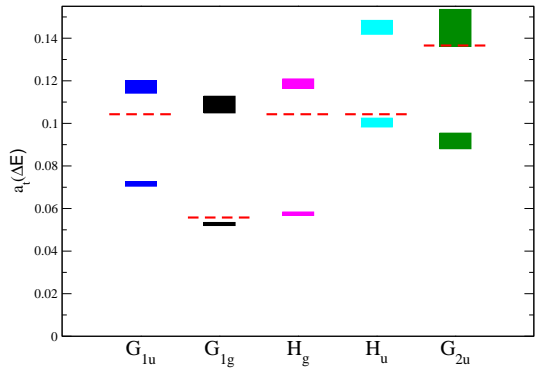}
   \caption{Splittings between the energy of the ground-state S-wave 
     static-light meson and other low-lying measured energies. 
     The dashed red lines show the scattering thresholds in each of
     the irreps. The position of the threshold in the $G_{1 g}$ irrep 
     corresponds to the rest mass of the pion. The thresholds in the 
     other irreps correspond to two-particle states containing a pion 
     with non-zero momentum.}
    \label{fig:thresholds}
\end{figure}

The analysis is complicated by the fact that 
on the $12^{3} \times 80$ lattice the smallest 
non-zero momentum component is 
approximately $0.088$, in units of $a_{t}^{-1}$, which lies between
the pion and rho masses. Therefore, the lowest-energy scattering
state in a given irrep may contain a rho meson at rest rather than a
pion with non-zero momentum. 
Moreover, the unit of momentum is also of the same order of magnitude  
as the splittings shown in Figure~\ref{fig:spectrum}, so it is possible that 
the threshold levels in some irreps correspond to states containing a static-light meson 
which lies in a representation other than $G_{1 u}$.    
We have considered both possibilities and find that on this spatial volume 
the threshold energies do in fact correspond to two-particle scattering states 
involving only the S-wave static-light meson and a pion. 
However, the threshold energy 
in $H_{g}$ is degenerate within errors with the energy of a scattering state consisting 
of an S-wave static-light meson and a rho meson at rest. 
The lowest threshold across the irreps is in $G_{1g}$ and it is associated with a pion at rest. 
The threshold energies in $G_{1 u}$, $H_{g}$ and $H_{u}$ correspond to 
a pion with momentum $(1,0,0)$ in units of $2\pi/a_{s}L$. In $G_{2 u}$ the lowest allowed pion momentum 
is $(1,1,0)$. 
In Figure~\ref{fig:thresholds} we have inserted 
the thresholds for multiparticle 
excitations into the splitting plot previously shown in Figure~\ref{fig:spectrum}.
The dashed red lines indicate the scattering thresholds, and  we see that multiparticle states  
should be detectable in all five irreps over the range of energies investigated in this 
study. 

%% file: discussion.tex
\section{Discussion}
\label{discussion}
Although a final determination of 
the excited state spectrum requires simulations at several 
lattice spacings, it is possible to draw some conclusions from 
the results obtained here. 

First we consider energy levels in $G_{1 u}$. 
The lowest-lying state in this irrep 
is the $J_{\ell}^{P_{\ell}} = \frac{1} {2}^{-}$ S wave. 
The next lowest angular momentum value allowed in 
this irrep is $J_{\ell} = \frac{7} {2}$. 
The $\frac{7} {2}^{-}$ state corresponds to an L=4 or 
G wave excitation. We expect such a state to be 
extremely heavy and also to appear as a degenerate level in other
lattice irreps. Since this is not the case we therefore conclude 
that the energy-levels lying directly above the ground state 
correspond to radial excitations of the $\frac{1} {2}^{-}$. 
Similarly, the lowest-energy state in the $G_{1 g}$ irrep is 
the $\frac{1} {2}^{+}$, and we expect the first excited state 
to be a radial excitation with the same quantum numbers. 
The interpretation of the higher-lying states is slightly more 
ambiguous due to the appearance of the $\frac{7} {2}^{+}$ F-wave 
state in this irrep.  
In the $H_{g}$ irrep the ground state has quantum numbers  
$J_{\ell}^{P_{\ell}} = \frac{3} {2}^{+}$. This state lies 
slightly above the $\frac{1} {2}^{+}$ ground state in the 
$G_{1 g}$ irrep. These are P-wave excitations. The data show
solid evidence that the natural ordering prevails in the P-wave sector
($m_{\frac{3}{2}+} > m_{\frac{1}{2}+}$). While this splitting is very
small it has been measured in this work to be different from zero at
the $5\sigma$ confidence level. The discussion of natural
ordering is made more complicated by the fact that the $G_{1g}$ ground
state lies close to the estimated S wave decay threshold.
The first excited state in $H_{g}$ lies just above its 
counterpart in $G_{1 g}$ and we also identify it as a radial 
excitation with P wave quantum numbers.

To make sense of the energy levels in 
$H_{u}$ one needs to consider them together with the 
states appearing in $G_{2 u}$. 
To begin with, we will assume that the variational analyses 
have succeeded in determining all the low-lying single particle 
energies in the $H_{u}$ and $G_{2 u}$ irreps, and consider 
how the relative ordering of D-wave states should effect the simulation results. 
Recall that both the $\frac{3} {2}^{-}$ and $\frac{5} {2}^{-}$ appear in the 
$H_{u}$ irrep and only the $\frac{5} {2}^{-}$ D-wave contributes to the 
$G_{2 u}$ irrep. In the standard ordering of states, the 
$\frac{3} {2}^{-}$ is lighter than the $\frac{5} {2}^{-}$ which implies 
that the lowest-lying state in $H_{u}$ should lie below the lowest-energy 
state in the $G_{2 u}$ irrep. One of the higher energy levels in the 
$H_{u}$ irrep, corresponding to the $\frac{5} {2}^{-}$ multiplet, 
should then be degenerate with the lowest energy level in $G_{2 u}$.  
This scenario is clearly not supported by the data.  

Another possibility is that inversion of the D-wave multiplets does occur.  
In this case, the ground state energies in $H_{u}$ and $G_{2 u}$ 
are degenerate up to lattice artifacts. The energies of the first excited 
states in these irreps overlap, and these states may correspond to a 
radial excitation of the $\frac{5} {2}^{-}$.  This would 
mean that the $\frac{3} {2}^{-}$ state is the third level in the $H_{u}$ 
irrep which has no counterpart in the $G_{2 u}$ irrep.  
However, it seems very unlikely that the splitting between the D-wave 
multiplets is so large. It could also be the case that systematic 
errors in the data are particularly severe in these channels, 
and that at finer lattice spacings and larger volumes the second energy 
level in $H_{u}$ shifts towards the ground state while the third energy 
level aligns itself with the second level in the $G_{2 u}$ irrep. 
One could then identify the second level in $H_{u}$ with the 
$\frac{3} {2}^{-}$ excitation, and the third level in $H_{u}$ and the 
second state in $G_{2 u}$ would correspond to a radial excitation 
of the $\frac{5} {2}^{-}$. However, given that the results were 
obtained using improved actions on a $(2.03~\rm{fm})^{3}$ spatial volume, 
it is improbable that systematic errors could account for such significant 
shifts in the spectrum. 

The inconsistencies between the data and the possible orderings 
of states seem to suggest that the assumption on which 
the preceding argument is based is incorrect, and we have 
not succeeded in measuring all of the lowest-lying single-particle 
energies in $H_{u}$ and $G_{2 u}$. 
In particular, there appear to be gaps in the spectrum 
of the $H_{u}$ irrep. This is not surprising when we recall that 
each variational basis consists of operators which have 
identical offsets. It appears that a complete determination of the spectrum will require 
a much greater variety of basis operators with varying degrees of overlap with 
all the low-lying states.    
Naively, the fact that the two lowest-lying measured energies in $H_{u}$ are close to their 
counterparts in $G_{2 u}$ might indicate that these are the energies of $\frac{5} {2}^{-}$ states.
However, we have 
already noted that the third level in $H_{u}$ does not have a partner in $G_{2 u}$, and there 
is no reason why the lower energies in $H_{u}$ could not correspond to $\frac{3} {2}^{-}$ excitations.   
Therefore, identification of the lowest measured energy levels in the 
$H_{u}$ irrep requires further investigation. 

In spite of the difficulty in interpreting the data, our 
study of the $H_{u}$ and $G_{2 u}$ irreps has been quite successful, and
we would like to emphasise two points. 
Firstly, we have been able to measure the energies of the ground-state  
$\frac{5} {2}^{-}$ D-wave and a radial excitation. 
These correspond to the first and second energy levels in the 
$G_{2 u}$ irrep. Secondly, the possibility that the interpolating operators 
are not coupling to all the low-energy states has only come to light 
because we have been able to precisely determine a number of energies in each 
of the irreps which we have compared to theoretical predictions.  

Finally, it is worth comparing our data with experimental data
for the $B_{s}$ spectrum. The energy difference between the
$B_{s J}^{*}(5850)$, with unknown quantum numbers, and
the ground-state pseudoscalar $B_{s}^{0}$ is
approximately 485~MeV. This number is very close to 
our value for the splitting between the ground-state S
wave and its first radial excitation. However, once
$1/m_{Q}$ corrections are taken into account, we see that
the splitting is also consistent with a P-wave identification
for the heavier state.

%% file: conclusions.tex
\section{Conclusions}
\label{conclusions}
We have presented an \textit{ab initio} study of 
the spectrum of static-light mesons in $N_{f}=2$ lattice 
QCD. Previously, computational constraints have made it difficult 
to determine the excited state spectrum to satisfactory accuracy. 
We have overcome these constraints using an efficient estimate for 
all elements of the lattice Dirac propagator. With this method and
exploiting translational 
invariance, we are able to average correlation functions over the 
whole volume of the lattice, yielding a dramatic reduction in the 
variance of our Monte Carlo estimates. 
The use of all-to-all propagators also facilitates the construction 
of spatially-extended interpolating operators which project 
onto the excited states. The construction of these operators and proper 
interpretation of the final numerical results depend on a clear 
understanding of the spatial symmetries of the lattice, and our 
results are given in terms of the irreducible representations 
of the octahedral point group. 

We have succeeded in determining a number of excited state splittings
in five of the six double-valued representations. 
A comparison of results obtained on two different spatial volumes indicates 
that that our results correspond to the energies of single-particle 
bound states. We have presented solid evidence (at the $5\sigma$
level) that the natural ordering of P-wave static-light mesons
prevails. We have been able to identify a number of radially excited states  
however a complete survey of the low-energy spectrum will 
require further work including, ultimately, an extrapolation to the 
continuum limit. 

A natural progression in this study would include the use 
of interpolating operators with more complicated offsets
which might better couple to heavier single-particle excitations, and the use of operators 
specifically constructed to couple to multiparticle states.    
We are also confident that a similar approach to the 
one used here can shed considerable light 
on the spectrum of heavy-light baryons. 

\begin{acknowledgments} 
This work was supported by the IITAC project, funded by the Irish
Higher Education Authority under PRTLI cycle 3 of the National
Development Plan and funded, in part, by an IRCSET postgraduate award.
We are grateful to the Trinity Centre for High-Performance Computing for their
support. We would like to thank Colin Morningstar and Jimmy Juge 
for kindly allowing us to use their correlator fit code. We also wish 
to thank Jon-Ivar Skullerud and Bu\u{g}ra Oktay for carefully reading this 
manuscript.
\end{acknowledgments} 